\pdfoutput=1

\documentclass[11pt]{article}
\usepackage{algpseudocode}
\usepackage{algorithm}

\usepackage[final]{acl}

\usepackage{times}
\usepackage{latexsym}
\usepackage{float}
\usepackage[T1]{fontenc}

\usepackage[utf8]{inputenc}
\usepackage{amsmath}
\usepackage{multirow}
\usepackage{microtype}

\usepackage{inconsolata}

\usepackage{xcolor}
\usepackage{graphicx}
\usepackage{subcaption}
\title{O\_O-VC: Synthetic Data-Driven One-to-One Alignment for Any-to-Any Voice Conversion}

\author{
 \textbf{Huu Tuong Tu\textsuperscript{1,2}}\quad
 \textbf{Huan Vu\textsuperscript{3}}\quad
 \textbf{Nguyen Tien Cuong\textsuperscript{1}}\quad
 \textbf{Ngo Dien Hy\textsuperscript{1,3}}\quad
\\
 \textbf{Nguyen Thi Thu Trang\textsuperscript{2$\thanks{Corresponding author}$}}
\\
 \textsuperscript{1}VNPT AI, VNPT Group \hspace*{2em} 
 \textsuperscript{2}Hanoi University of Science and Technology
 \\
 \textsuperscript{3}Business AI Lab, National Economics University
 \\
 \textsuperscript{huutu12312vn@gmail.com, huanv@neu.edu.vn, nguyentiencuong@vnpt.vn, ngodienhy@vnpt.vn, trangntt@soict.hust.edu.vn}
}

\begin{document}
\maketitle
\begin{abstract}
 Traditional voice conversion (VC) methods typically attempt to separate speaker identity and linguistic information into distinct representations, which are then combined to reconstruct the audio. However, effectively disentangling these factors remains challenging, often leading to information loss during training. In this paper, we propose a new approach that leverages synthetic speech data generated by a high-quality, pretrained multispeaker text-to-speech (TTS) model. Specifically, synthetic data pairs that share the same linguistic content but differ in speaker identity are used as input-output pairs to train the voice conversion model. This enables the model to learn a direct mapping between source and target voices, effectively capturing speaker-specific characteristics while preserving linguistic content. Additionally, we introduce a flexible training strategy for any-to-any voice conversion that generalizes well to unseen speakers and new languages, enhancing adaptability and performance in zero-shot scenarios. Our experiments show that our proposed method achieves a 16.35\% relative reduction in word error rate and a 5.91\% improvement in speaker cosine similarity, outperforming several state-of-the-art methods. Voice conversion samples can be accessed at: \url{https://oovc-emnlp-2025.github.io/}
\end{abstract}

\section{Introduction}
\label{sec:introduction}

Voice conversion specifically aims to transform a source speaker’s voice to match a target speaker while preserving the original linguistic content. This is typically done by disentangling speech into content and speaker identity representations, which are combined during training to reconstruct the audio. At inference time, the source content is paired with a target speaker embedding to generate the converted speech.

Several methods have been proposed for VC, with supervised training being a common approach. Content encoders are trained with text labels to extract linguistic features, and speaker encoders use speaker labels to capture identity-specific traits \citep{w2vc, ppg1}. Alternatively, phonetic posteriorgrams (PPGs) can be used directly as content representations \citep{ppg, ppg2}. However, both approaches often struggle to capture speaker-independent prosody and accent information. Moreover, training content encoders as ASR models can introduce alignment errors or recognition errors, which can negatively impact conversion quality \citep{ace-vc}.

On the other hand, some methods avoid using text labels by leveraging self-supervised learning (SSL) to extract high-level phonetic representations \citep{ssl1, ssl2, ssl3, ssl4}. These approaches aim to remove speaker identity from source audio while preserving speaker-independent features such as accent and content. To achieve this separation, techniques such as vector quantization \citep{vqvc}, instance normalization \citep{ain}, heuristic transformation \citep{selfvc}, bottleneck, and data augmentation \citep{free-vc} are commonly applied. However, despite these efforts, such methods still struggle to completely eliminate speaker information from the source speech. This often leads to speaker leakage, where the converted audio retains unintended characteristics of the source speaker, resulting in mismatches between the synthesized voice and the intended target speaker \citep{knn-vc}.

Previous studies have primarily focused on feature disentanglement methods and audio reconstruction in voice conversion systems. However, feature disentanglement remains a challenging task and training models to reconstruct the audio may not be well suited for the voice conversion objective, which inherently involves transforming speech from one speaker to another. To address these limitations, we propose a novel training strategy that leverages synthetic data generated by a high-quality multi-speaker text-to-speech (TTS) system to directly establish input-output mappings for voice conversion, bypassing the need for traditional reconstruction-based approaches. Despite the high fidelity of synthetic data, such TTS systems are typically constrained to a fixed set of speakers, limiting their applicability to any-to-any voice conversion, particularly for unseen speakers. To mitigate this issue, we introduce a training framework that promotes generalization to unseen speakers without relying on additional text or speaker labels, thereby enhancing the system’s adaptability and performance in zero-shot voice conversion scenarios. In summary, we make the following contributions:
\begin{itemize}
    \item Synthetic data for voice conversion training: We propose the use of synthetic data generated by a high-quality multi-speaker TTS system to train voice conversion models. This approach eliminates the need for audio reconstruction and feature disentanglement, enabling direct learning of input-output mappings.
    \item Improved generalization: We introduce a training strategy that allows the model to generalize to unseen speakers or unseen languages, making it well suited for zero-shot voice conversion.
    \item We validate the effectiveness of our approach through extensive experiments, showing significant improvements over traditional reconstruction-based methods, especially in challenging zero-shot settings. 
\end{itemize}

\section{Literature Review}
\label{sec:background}
The goal of voice conversion (VC) is to transform the voice of a source speaker into that of a target speaker while preserving the original linguistic content. Achieving this requires an effective decomposition of speech signals into distinct components such as linguistic content, speaker timbre, and prosodic characteristics. Early VC systems were typically trained as speech-to-speech models on parallel datasets, where multiple speakers uttered the same sentences, defining the task as a sequence-to-sequence problem \citep{parallel3, parallel4}. Recent VC approaches have focused on reconstructing speech using disentangled representations of linguistic content and speaker identity.

\subsection{Text-Based Method}

A common strategy is to leverage pretrained automatic speech recognition (ASR) models to extract phonetic features such as PPGs, which provide a speaker-independent representation of the input speech \citep{ppg, ppg1, ppg2}. Specifically, speaker information is obtained using a pretrained speaker verification (SV) model.  The speaker embeddings are combined with the content features during decoding, allowing the system to generate speech in the target speaker’s voice. Some approaches leverage hidden text representations from pretrained multispeaker text-to-speech (TTS) models, using them either as semantic features or as target representations for learning a mapping from audio to text \citep{transfertts, transfertts1}.

Despite their advantages, text-based methods suffer from several limitations. PPGs and other textual representations often fail to capture fine-grained attributes such as accent, prosody, and speaker-independent speaking style. As a result, these systems often produce speech that lacks expressiveness and sounds overly neutral \citep{ace-vc}. Although ASR-based disentanglement methods have shown progress in separating speaker and content information, their reliance on textual supervision and limited prosodic modeling remain significant challenges for achieving natural and expressive voice conversion across diverse speakers and languages.

\subsection{Text-Free Method}
To address the limitations of text-based methods, text-free approaches have emerged, leveraging self-supervised learning models to extract content representations without requiring transcriptions \citep{ssl1, ssl2, ssl3, ssl4}. Although self-supervised learning (SSL) features capture high-level information related to linguistic content, they often retain residual speaker characteristics. To address this, methods such as bottleneck layers \citep{free-vc}, vector quantization \citep{vqvc}, and instance normalization \citep{ain} have been proposed to compress SSL features and extract speaker-independent content representations. However, effective disentanglement heavily depends on the choice of bottleneck configuration: if the bottleneck dimension is too large, speaker information may be retained; if too small, important content information can be lost. A similar trade-off exists in vector quantization: large codebooks may retain speaker traits, while overly small codebooks may lead to excessive loss of content information. Moreover, these compression techniques often degrade the quality of the generated audio, and full disentanglement of speaker and content information remains an open challenge.

\subsection{KNN Method}
Recent work has introduced k-nearest neighbor (kNN)-based voice conversion methods \citep{knn-vc}, offering a simpler alternative to traditional feature disentanglement approaches. These methods operate directly on frame-level self-supervised representations extracted from both source and target speech, which encode both phonetic and speaker-specific information. Voice conversion is performed by replacing each frame of the source with its nearest neighbor from the target set, followed by vocoder-based synthesis.

However, in one-shot scenarios, the limited size of the target set restricts the pool of candidate neighbors, often resulting in higher word error rates. To address this, the Phoneme Hallucinator \citep{hallucinator} was proposed, leveraging a permutation network to synthesize additional target representations and expand the neighbor set, thereby improving intelligibility. Nonetheless, averaging features in kNN-based retrieval can lead to oversmoothing, reducing speaker distinctiveness and clarity in the synthesized speech.

\subsection{Diffusion Method}
Diffusion models have shown exceptional performance in generative tasks across a variety of domains, including images, videos, and audio. In speech processing, diffusion models have been successfully applied to tasks such as audio generation \citep{ag_diff1, ag_diff2} and text-to-speech (TTS) synthesis \citep{ttsdiff1, ttsdiff2}. Furthermore, diffusion models have been investigated for VC tasks with the aim of enhancing the conversion process. In particular, diffusion-based VC models \citep{diffvc} have demonstrated high performance in zero-shot speaker adaptation through iterative sampling processes. While recent works such as Diff-HierVC \citep{diffhiervc} and DDDM-VC \citep{dddmvc} have further improved zero-shot VC performance through source-filter disentanglement and disentangled denoising processes, the audio quality of diffusion models is still limited.

\section{Methodology}
\label{sec:proposed}

In text-free VC systems, content and speaker identity are often not fully disentangled. As a result, speaker information can leak into the content representation, which undermines the system’s ability to perform clean speaker conversion. This leakage reduces the system’s generalization to unseen voices or speaking styles and often results in converted speech that retains characteristics of the source speaker. 

In text-based VC, ASR-derived content representations are highly sensitive to transcription errors, mispronunciations, and noisy labels, which can compromise their reliability and degrade the quality of converted speech \citep{ppg, ppg1, ppg2}. Some approaches attempt to leverage knowledge transfer from hidden representations of text encoders in multispeaker TTS models \citep{transfertts, transfertts1}. However, mapping audio representations directly to these text-based features is a difficult task. In addition, this process typically requires an explicit alignment mechanism between speech and text, which introduces further complexity.

As an alternative solution, synthetic data offers several advantages for voice conversion. When both source and target audio are generated from the same linguistic content, it provides a clean and direct supervisory signal. This shared content allows precise frame-level alignment between source and target audio, enabling more stable and fine-grained learning of the conversion function. Unlike traditional approaches that rely on symbolic representations (e.g., phonemes or characters), synthetic data eliminates the need for such intermediates and avoids issues like label noise commonly found in real-data training. Furthermore, with controlled or predefined durations, speaker-independent features are inherently aligned across domains, removing the need for forced alignment algorithms used in previous work \citep{transfertts, transfertts1}. Unlike methods that map audio to hidden text representations from multispeaker TTS models, our approach directly maps source audio to target audio, simplifying the learning process. This enables one-to-one frame alignment, allowing the model to focus more effectively on speaker transformation while preserving linguistic content. Finally, synthetic data enables the creation of diverse speaker pairs with uniform content, supporting the learning of generalizable speaker conversion mappings. Motivated by these advantages, this work is the first to propose using synthetic data as a training paradigm for voice conversion models.

\subsection{Synthetic Data Strategy}
\label{subsec:synthetic-strategy}
Building upon these advantages, we propose a synthetic data strategy to improve the disentanglement of speaker and content representations in voice conversion. Instead of relying on real-world utterances, we generate high-quality synthetic pairs with identical linguistic content but varying speaker identities. This provides ideal supervision for isolating speaker-independent content.

We use a multi-speaker TTS system that produces natural, intelligible speech across speakers from a shared linguistic latent space. Our selection criteria for the TTS system are: (1) the generated speech must be of high fidelity, exhibiting natural prosody and clarity, and (2) the model must synthesize source and target utterances from the same linguistic latent space, ensuring consistent phonetic and prosodic alignment across speakers. These criteria ensure that the synthetic speech pairs are perfectly aligned in linguistic structure while differing only in speaker identity.

We adopt VITS \citep{vits} as the backbone of our TTS system because it combines variational inference, flows, and adversarial learning to generate high-quality speech with precise duration control and the ability to sample two audios with different speakers conditioned on a shared latent linguistic representation. This enables the creation of large-scale, controllable training data that improves disentanglement, reduces speaker leakage, and enhances voice conversion robustness.

Given a text input \( c_{\text{text}} \), source speaker embedding \( s_{\text{src}} \), target speaker embedding \( s_{\text{tgt}} \), and noise vector \( w \), we generate a pair of utterances by first encoding the linguistic content:
\begin{equation}
h_{\text{text}} = \text{TextEncoder}(c_{\text{text}})
\end{equation}
We then sample a global speaker token:
\begin{equation}
g = \text{random}(s_{\text{src}}, s_{\text{tgt}})
\end{equation}
and predict the duration based on speaker condition using the duration predictor (DP):
\begin{equation}
\text{dur}_{\text{text}} = \text{DP}(h_{\text{text}}, g, w)
\end{equation}
Next, we project the encoded content into a linguistic latent distribution:
\begin{equation}
\mu_p, \sigma_p = \text{Projector}(h_{\text{text}})
\end{equation}
We then sample latent linguistic features by applying duration expansion through the length regulator (LR), defined as follows:
\begin{equation}
\mu_p, \sigma_p = \text{LR}(\mu_p, \sigma_p, \text{dur}_{\text{text}})
\end{equation}
\begin{equation}
z_p = \mu_p + \sigma_p \cdot \epsilon, \quad \epsilon \sim \mathcal{N}(0, I)
\end{equation}
To generate speaker-specific representations, we apply the inverse flow conditioned on each speaker:
\begin{equation}
z_{\text{src}} = \text{Flow}^{-1}(z_p, s_{\text{src}})
\end{equation}
\begin{equation}
z_{\text{tgt}} = \text{Flow}^{-1}(z_p, s_{\text{tgt}})
\end{equation}

\begin{figure*}[t]
  \centering
  \begin{subcaptionbox}{Training\label{fig:training}}[0.5\linewidth]
    {\includegraphics[width=1.1\linewidth]{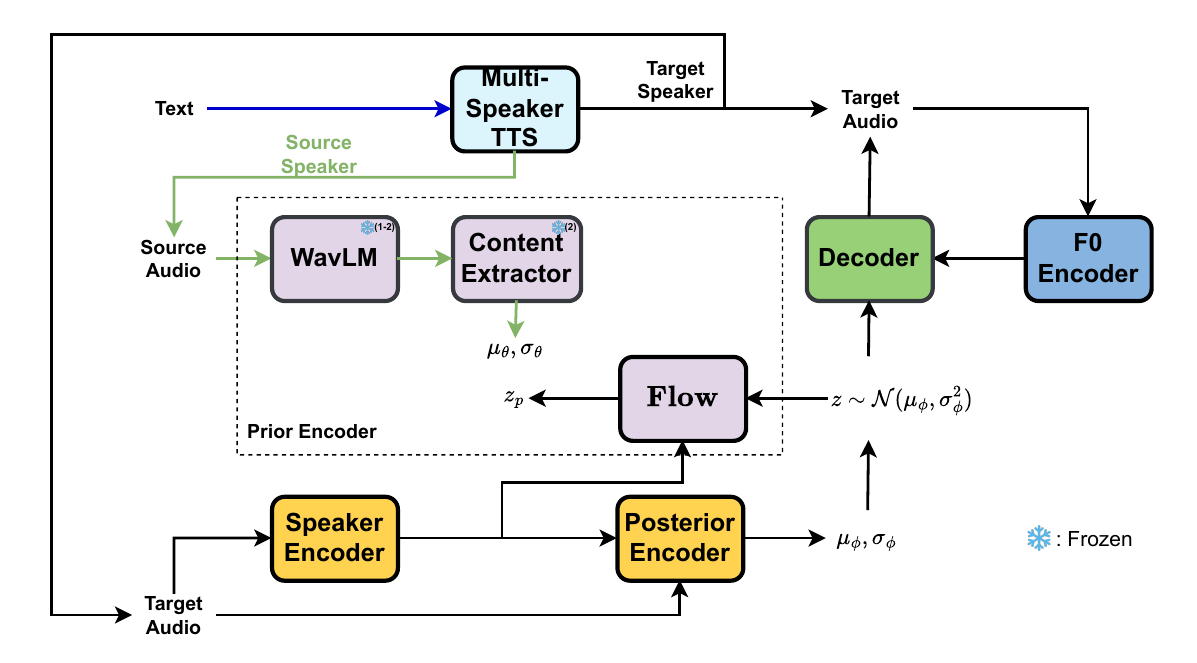}}
  \end{subcaptionbox}
  \hfill
  \begin{subcaptionbox}{Inference\label{fig:inference}}[0.45\linewidth]
    {\raisebox{6mm}{\includegraphics[width=1.05\linewidth]{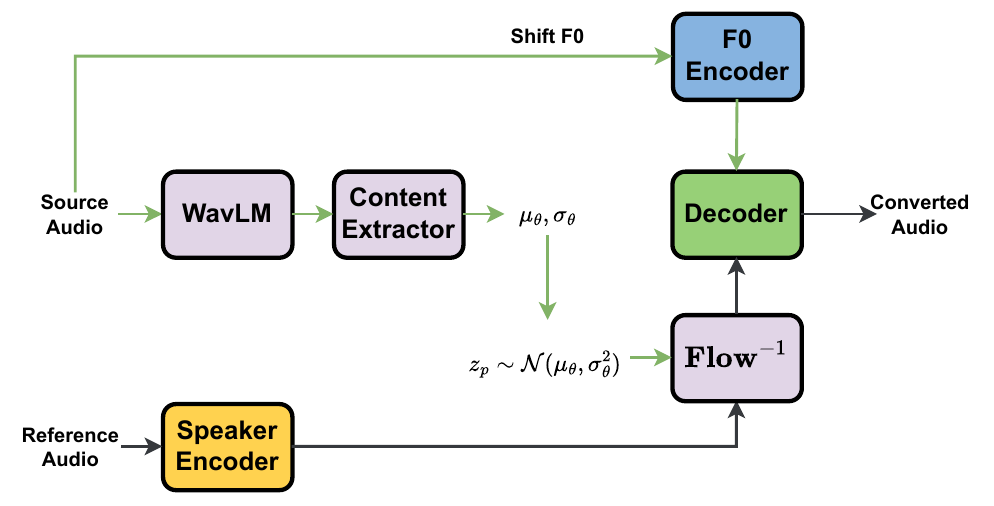}}}
  \end{subcaptionbox}
  \caption{Voice conversion with synthetic data.}
  \label{fig:train_infer}
\end{figure*}

Finally, we decode both representations into waveform audio:
\begin{equation}
a_{\text{src}} = \text{Decoder}(z_{\text{src}}), \quad a_{\text{tgt}} = \text{Decoder}(z_{\text{tgt}})
\end{equation}

This process generates pairs of utterances that share the same linguistic latent features, duration, and prosody, while differing only in speaker identity. Such pairs provide a clean and consistent training signal for voice conversion. Standard VITS, however, produces utterances independently, which often leads to mismatches in duration and rhythm.

\subsection{Model Overview}

After generating synthetic data consisting of source and target speech pairs for supervised training, these utterances are directly utilized as input-output pairs for the voice conversion model. As the backbone architecture, we adopt a VITS-base model. Following the design of FreeVC \citep{free-vc}, our model structure retains its core components. However, we note that while the source and target speech pairs share the same underlying linguistic content, the target speech is conditioned on a different speaker identity, which primarily manifests in variations in pitch. To avoid mismatch between input and output during training, we incorporate the fundamental frequency ($F0$) of the target speech as an additional conditioning feature when decoding the final audio. The general model pipeline is illustrated in Figure \ref{fig:train_infer}.

\subsubsection{Training Procedure}

In the training phase, the source and target audios are processed through different stages:

\noindent\textbf{Source Audio Processing:} The source audio is passed through a pretrained WavLM \citep{wavlm} and a content extractor to obtain a distribution of content features $\mathcal{N}(\mu_{\theta}, \sigma^2_{\theta})$.

\noindent\textbf{Target Audio Processing:} The target audio is passed through a speaker encoder and a posterior encoder to extract the posterior latent distribution $\mathcal{N}(\mu_{\phi}, \sigma^2_{\phi})$. Then, a latent variable $z$ is sampled from this distribution
$z \sim \mathcal{N}(\mu_{\phi}, \sigma^2_{\phi})$.
The sampled latent vector $z$ is passed through a flow-based module to obtain $z_p$, transforming the posterior distribution to match the prior distribution. A Kullback-Leibler (KL) divergence loss is calculated to minimize the discrepancy between the posterior and prior distributions.

\noindent\textbf{Fundamental Frequency ($\textbf{F0}$) Adjustment:} To address the mismatch in $F0$ between the source and target audio, we extract the $F0$ of the target audio and pass it through an $F0$ encoder to obtain pitch-related features. The decoder then takes the transformed latent representation along with the $F0$ features to generate the target audio. In this work, we extract $F0$ using Parselmouth\footnote{\url{https://github.com/YannickJadoul/Parselmouth}}.

\subsubsection{Fine-Tuning Adaptation}

Training with synthetic data often results in poor generalization to unseen speakers. Multi-speaker TTS models are typically trained with fixed speaker embeddings, which can lead to reduced speaker similarity for out-of-domain speakers. Moreover, slight discrepancies in the linguistic content between source and target utterances may persist, despite both being generated from the same text.

To mitigate these challenges, we adopt a two-phase training strategy:

\begin{itemize}
    \item \textbf{Phase 1:} Train the model with synthetic data, where the WavLM and content extractor components are responsible for learning independent speaker representations. The WavLM is frozen during phase 1. 
    \item \textbf{Phase 2:} Fine-tune the model using real large multispeaker speech recordings corpus with a reconstruction-based objective, the input and output of the model is the same audio. During this phase, WavLM and content extractor is frozen to preserve speaker-independent representations learned in phase 1. This fine-tuning phase helps the model adapt to new speakers without the need for transcripts and improves the fidelity of linguistic content. Additionally, the model becomes more versatile and can easily adapt to different domains, such as various languages or accents.
\end{itemize}

\subsubsection{Inference}
During inference, the source audio is processed by WavLM and the content extractor to obtain the content distribution $\mathcal{N}(\mu_{\theta}, \sigma^2_{\theta})$. A latent sample $z_p$ is drawn and combined with the speaker embedding from the reference audio via the inverse flow model, producing a feature that captures both content and speaker information.

To address the mismatch in $F0$ between the source and reference audio, we shift $F0$ source to $F0$ target with same median level, the following steps are performed:
\begin{align}
F0_{\text{src}} &= \text{Get\_F0}(\text{Source Audio}) \\
F0_{\text{ref}} &= \text{Get\_F0}(\text{Reference Audio}) \\
\log F0_{\text{shifted}} &= \log (F0_{\text{src}}) - \text{med}(\log (F0_{\text{src}})) \notag\\
&\quad + \text{med}(\log (F0_{\text{ref}}))\\
F0_{\text{shifted}} &= \exp(\log F0_{\text{shifted}})
\end{align}
The shifted $F0$ is then used as a conditioning feature along with the fused linguistic and speaker representation to generate the final audio output.

\subsubsection{Objective Function}
Following the methodology proposed in \citep{free-vc}, we formulate the objective function by combining losses from conditional variational autoencoders (CVAE) and generative adversarial networks (GAN) \citep{gan,fm}. CVAE-related losses include the KL divergence loss $L_{\text{kl}}$, which measures the discrepancy between the prior and posterior distributions of the flow-based model, and a phase-dependent reconstruction/conversion loss, either $L_{\text{rec}}$ in phase 2 or $L_{\text{cv}}$ in phase 1, defined as the L1 distance between the predicted and target mel-spectrograms. GAN-related losses include the adversarial loss for the discriminator $L_{\text{adv}}(D)$, the adversarial loss for the generator $L_{\text{adv}}(G)$, and the feature matching loss $L_{\text{fm}}(G)$. We further incorporate two distillation losses into the total objective. The final loss function is defined as:
\begin{equation}
    L(D) = L_{\text{adv}}(D)
\end{equation}
\begin{equation}    
    L(G) = L_{\text{rec/cv}} + L_{\text{kl}} + L_{\text{adv}}(G) + L_{\text{fm}}(G)
\end{equation}

\begin{table*}
\begin{center}

\scalebox{0.96}{
\begin{tabular}{|l|cccc|ccc|}
\hline
\multicolumn{1}{|c|}{\multirow{2}{*}{\textbf{Model}}} & \multicolumn{4}{c|}{\textbf{Objective Evaluation}}                                 & \multicolumn{3}{c|}{\textbf{Subjective Evaluation}}                                        \\ \cline{2-8} 
\multicolumn{1}{|c|}{}                                & \textbf{SECS$\uparrow$}       & \textbf{WER$\downarrow$}       & \textbf{CER$\downarrow$}       & \textbf{NISQA$\uparrow$}     & \textbf{MOS$\uparrow$}                      & \textbf{SMOS$\uparrow$}                     & \textbf{B-MOS$\uparrow$}     \\ \hline
FreeVC                                                & 75.66               & 2.37               & 0.78               & \textbf{\textcolor{blue}{4.60}}      & $\textbf{\textcolor{blue}{3.60}} \pm \textbf{\textcolor{blue}{0.26}}$ & $3.01 \pm 0.28$                   & $\underline{3.31}$ \\ \hline
KNN-VC                                                & 78.33               & 2.16               & $\underline{0.62}$ & 3.92               & $3.17 \pm 0.23$                   & $2.89 \pm 0.21$                   & 3.03               \\ \hline
Diff-Hier                                             & 81.42               & 3.82               & 1.51               & 3.80               & $2.87 \pm 0.28$                   & $3.42 \pm 0.25$                   & 3.15               \\ \hline
DDDM-VC                                               & $\underline{81.86}$ & 6.84               & 2.92               & 3.91               & $2.89 \pm 0.28$                   & $\textbf{\textcolor{blue}{3.61}} \pm \textbf{\textcolor{blue}{0.23}}$ & 3.25               \\ \hline
Facodec                                               & 81.54               & $\underline{2.08}$ & 0.64               & 3.90               & $2.49 \pm 0.29$                   & $2.66 \pm 0.27$                   & 2.58               \\ \hline
O\_O-VC (Ours)                                               & \textbf{\textcolor{blue}{86.70}}      & \textbf{\textcolor{blue}{1.74}}      & \textbf{\textcolor{blue}{0.53}}      & $\underline{4.04}$ & $\underline{3.42 \pm 0.24}$       & $\underline{3.48 \pm 0.23}$       & $\textbf{\textcolor{blue}{3.45}}$    \\ \hline
\end{tabular}
}
\caption{Any-to-any voice conversion results. \textbf{\textcolor{blue}{Blue}} indicates the best performance, \underline{Underline} indicates second best. Subjective evaluation results showing MOS and SMOS scores, along with 95\% confidence intervals.
}
\label{table:result}
\end{center}
\end{table*}

\section{Experiment Setup}
\subsection{Datasets}
For phase 1 of training, we use synthetic speech generated by a publicly available pretrained VITS model\footnote{\url{https://github.com/jaywalnut310/vits}}. Specifically, we adopt the model released in the official repository. The amount of synthetic data corresponds to the VCTK training set used in the original VITS implementation. For each sample, the target audio is synthesized using the ground-truth text and speaker ID, while the source audio is generated by sampling a different random speaker ID. In phase 2, we fine-tune the model on the LibriSpeech dataset \citep{librispeech}, using the train-clean-360 and train-clean-100 subsets, totaling approximately 460 hours of speech from 1,172 speakers. Evaluation is conducted on the test-clean subset under any-to-any voice conversion scenarios.

\subsection{Model Configuration and Training Details}

We follow the implementation and hyperparameter setup of FreeVC \citep{free-vc}. Training occurs in two phases: up to 450k steps on synthetic data, followed by 150k steps of fine-tuning on real speech. All experiments are conducted on four NVIDIA Tesla A100 GPUs.

We compare our method with several recent state-of-the-art voice conversion models, including FreeVC \citep{free-vc}, DDDM-VC \citep{dddmvc}, Diff-HierVC \citep{diffhiervc}, FaCodec (NaturalSpeech 3) \citep{ns3}, and KNN-VC \citep{knn-vc}. For all baselines, we use official publicly released pretrained models. For KNN-VC, we use an 8-minute real speech segment as the reference pool for nearest-neighbor retrieval.

\subsection{Evaluation Metrics}
\label{sec:evaluate-metrics}
\noindent\textbf{Objective Evaluation}: We evaluate system performance using four objective metrics: Character Error Rate (CER), Word Error Rate (WER), Speaker Encoder Cosine Similarity (SECS), and Objective Naturalness. CER and WER assess intelligibility between the source and converted speech, using the HuBERT model\footnote{\url{https://huggingface.co/facebook/hubert-large-ls960-ft}} \citep{hubert}. SECS measures speaker similarity using the cosine similarity between embeddings extracted by Resemblyzer\footnote{\url{https://github.com/resemble-ai/Resemblyzer}}. Naturalness is assessed using NISQA \citep{nisqa}, which estimates perceptual speech quality without reference audio. We compute these metrics on 1,000 randomly sampled audio pairs from LibriSpeech test-clean.

\noindent\textbf{Subjective Evaluation}: For human evaluation, we use Mean Opinion Score (MOS) and Speaker Similarity Mean Opinion Score (SMOS). MOS rates naturalness, while SMOS rates speaker similarity, both on a 1-5 scale. We randomly select 30 audio pairs from the objective set, each evaluated by three different annotators, resulting in a total of 540 labeled audio samples. A total of 12 volunteer listeners participate in the evaluation. Final scores are calculated by averaging the ratings across annotators for each pair to ensure reliability. In voice conversion systems, both MOS and SMOS help evaluate model quality. To provide an overall comparison, we introduce a new metric called balance-MOS (B-MOS), defined as the average of these two scores.

\begin{table}[!htbp]
\centering
\scalebox{0.85}{
\begin{tabular}{|l|c|c|c|c|}
\hline
\textbf{Model}                    & \textbf{SECS$\uparrow$}       & \textbf{WER$\downarrow$}         & \textbf{CER$\downarrow$}         & \textbf{NISQA$\uparrow$}       \\ \hline
O\_O-VC (Ours)                     & \underline{86.70}   & \textbf{\textcolor{blue}{1.74}}        & \textbf{\textcolor{blue}{0.53}}        & 4.04                 \\ \hline
w/o F0 Encoder                   & \textbf{\textcolor{blue}{87.00}}      & \underline{2.07}     & \underline{0.61}     & 3.85                 \\ \hline
w/o Finetuning                   & 70.78               & 2.18                 & 0.66                 & \underline{4.59}     \\ \hline
FreeVC                & 75.66               & 2.37                 & 0.78                 & \textbf{\textcolor{blue}{4.60}}        \\ \hline
\end{tabular}
}
\caption{Ablation study results.}
\label{result:ablation}
\end{table}

\section{Results and Analysis}

\begin{figure*}[htbp]
    \centering
    \hspace{0.05\textwidth}
        \includegraphics[width=0.95\linewidth]{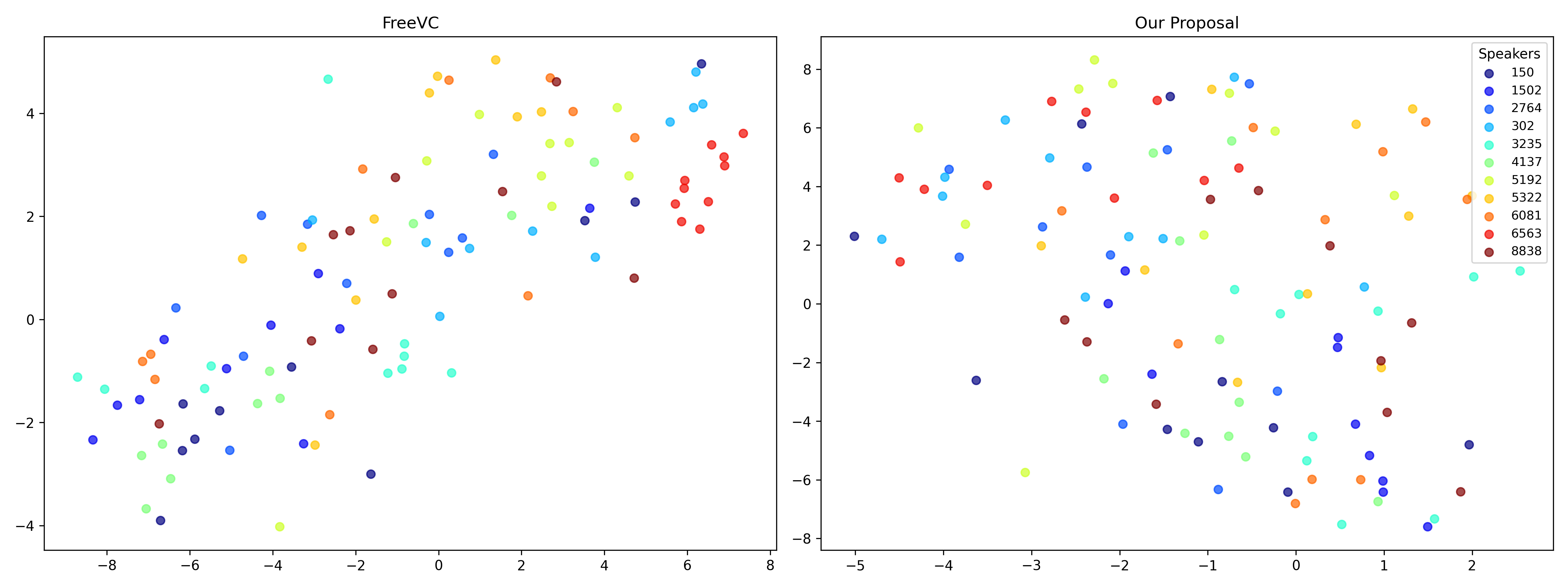}
    \caption{T-SNE visualization of speaker-independent features. More distributed points with no clusters indicate better speaker independence.}
    \label{fig:comparison}
\end{figure*}

\subsection{Zero-Shot Voice Conversion} 
We evaluate our model in a zero-shot setting, where the target speaker is unseen during training. The results in Table \ref{table:result} demonstrate that our model achieves the best performance in terms of content consistency, with the lowest WER and CER. Furthermore, our model achieves the second highest MOS, only slightly behind FreeVC. This can be attributed to the fact that FreeVC is trained on a high-quality speech dataset, whereas our model is fine-tuned and adapted on LibriSpeech, which is of comparatively lower quality, leading to decreased performance. Despite FreeVC's strong MOS, it performs notably worse in terms of speaker similarity and content intelligibility compared to our model. Although DDDM-VC achieves the highest Similarity Mean Opinion Score (SMOS), its speech quality is comparatively poor. Overall, our model achieves the best intelligibility while maintaining a strong balance between naturalness (MOS) and speaker similarity (SMOS), outperforming recent systems in a zero-shot scenario.


\begin{figure}[!htbp]
    \centering
    \hspace{0.05\textwidth}
        \includegraphics[width=0.95\linewidth]{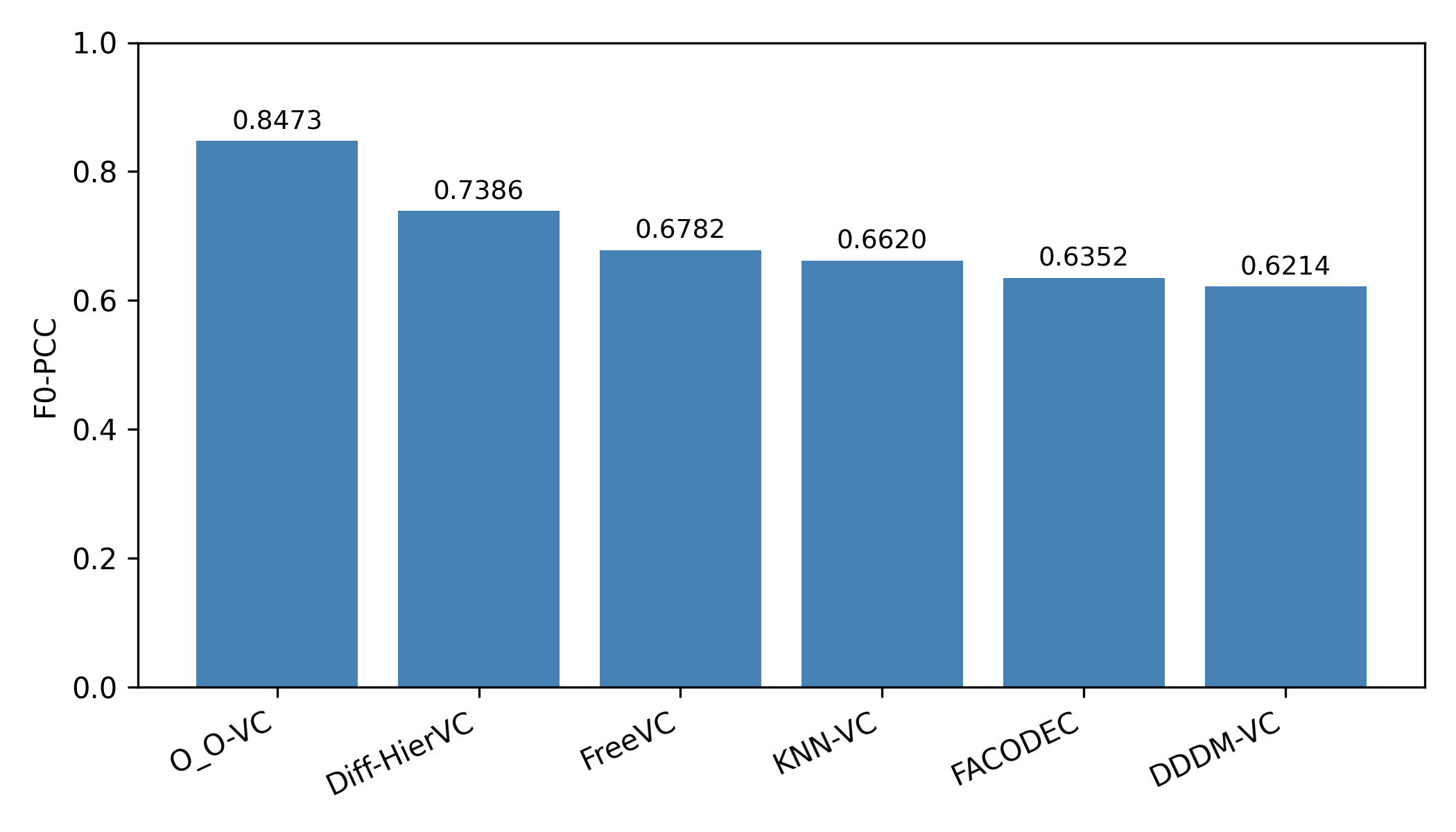}
    \caption{Comparison of systems on F0-PCC}
    \label{figure:f0-pcc}
\end{figure}

\subsection{Ablation Study} 
\label{sec:ablation}
We conduct an ablation study by modifying or removing key modules to evaluate their individual contributions, summarized in Table \ref{result:ablation}. We observe that removing the use of synthetic data, the F0 encoder, or phase 2 fine-tuning each leads to a noticeable drop in intelligibility, highlighting the importance of all three components. Eliminating phase 2 fine-tuning also causes a significant reduction in speaker similarity, likely due to the limited speaker diversity in the phase 1 dataset. However, since the phase 1 data is of higher quality, the phase 2 adaptation may slightly reduce speech quality. These findings demonstrate that using a synthetic dataset in phase 1 can achieve speech quality comparable to real data (competitive NISQA with FreeVC), while our phase 2 adaptation enables the model to generalize effectively to new datasets and unseen speakers without transcript labels.

In addition to intelligibility and naturalness, we also assess pitch preservation by reporting the F0 Pearson Correlation Coefficient (F0-PCC) \citep{pcc}, which is computed between the F0 contours of the source and converted audio. As shown in Figure~\ref{figure:f0-pcc}, our system achieves the highest F0-PCC, outperforming recent models that explicitly condition on F0 such as Diff-HierVC, as well as the same backbone model FreeVC without F0 conditioning. These results highlight the effectiveness of the proposed F0 encoder in maintaining accurate pitch contours and demonstrate strong pitch consistency across converted speech.

We also quantitatively evaluate how effectively the prior encoder removes speaker information by comparing our model to FreeVC, which shares the same backbone architecture. Our goal is to demonstrate that training with synthetic data significantly improves the removal of speaker identity from source audio. To assess this, we use three clustering evaluation metrics: Adjusted Rand Index (ARI), Normalized Mutual Information (NMI) and Silhouette Score. The ARI measures the similarity between predicted clusters and true speaker labels, adjusted for chance. A lower ARI indicates that the clusters do not correspond well to speaker identities, suggesting better speaker information removal. NMI measures the amount of shared information between the predicted and true clusters; lower values indicate weaker correlation and thus stronger speaker anonymization. The Silhouette Score reflects how well each embedding fits within its cluster compared to the others. Lower scores imply that the model's embeddings are not tightly grouped by speaker, further indicating that speaker identity has been suppressed. The quantitative results are shown in Table \ref{result:cluster}. Our model, trained with synthetic data, consistently achieves lower scores across all metrics, demonstrating its improved ability to remove speaker-specific information compared to FreeVC.

\begin{table}[]
\begin{tabular}{|l|c|c|c|}
\hline
\multicolumn{1}{|c|}{\textbf{Model}} & \textbf{ARI$\downarrow$}  & \textbf{NMI$\downarrow$}  & \textbf{Silhouette$\downarrow$} \\ \hline
O\_O-VC (Ours)                         & \textbf{\textcolor{blue}{0.07}} & \textbf{\textcolor{blue}{0.31}} & \textbf{\textcolor{blue}{0.15}}       \\ \hline
FreeVC                               & 0.13          & 0.41          & 0.17                \\ \hline
\end{tabular}
\caption{Evaluation of speaker information removal.}
\label{result:cluster}

\end{table}

For intuitive visualization, we use t-SNE to plot the speaker-independent features in Figure \ref{fig:comparison}. Our model's embeddings are more evenly dispersed across speakers (different colors), indicating greater speaker independence. In contrast, FreeVC shows noticeable clustering, such as for speakers $6563$ and $5192$, which indicates that its features preserve more speaker-specific information.

\subsection{Adaptation to New Languages} 
\label{sec:adapt}
\begin{figure}[!htbp]
    \centering
    \hspace{0.05\textwidth}
        \includegraphics[width=0.95\linewidth]{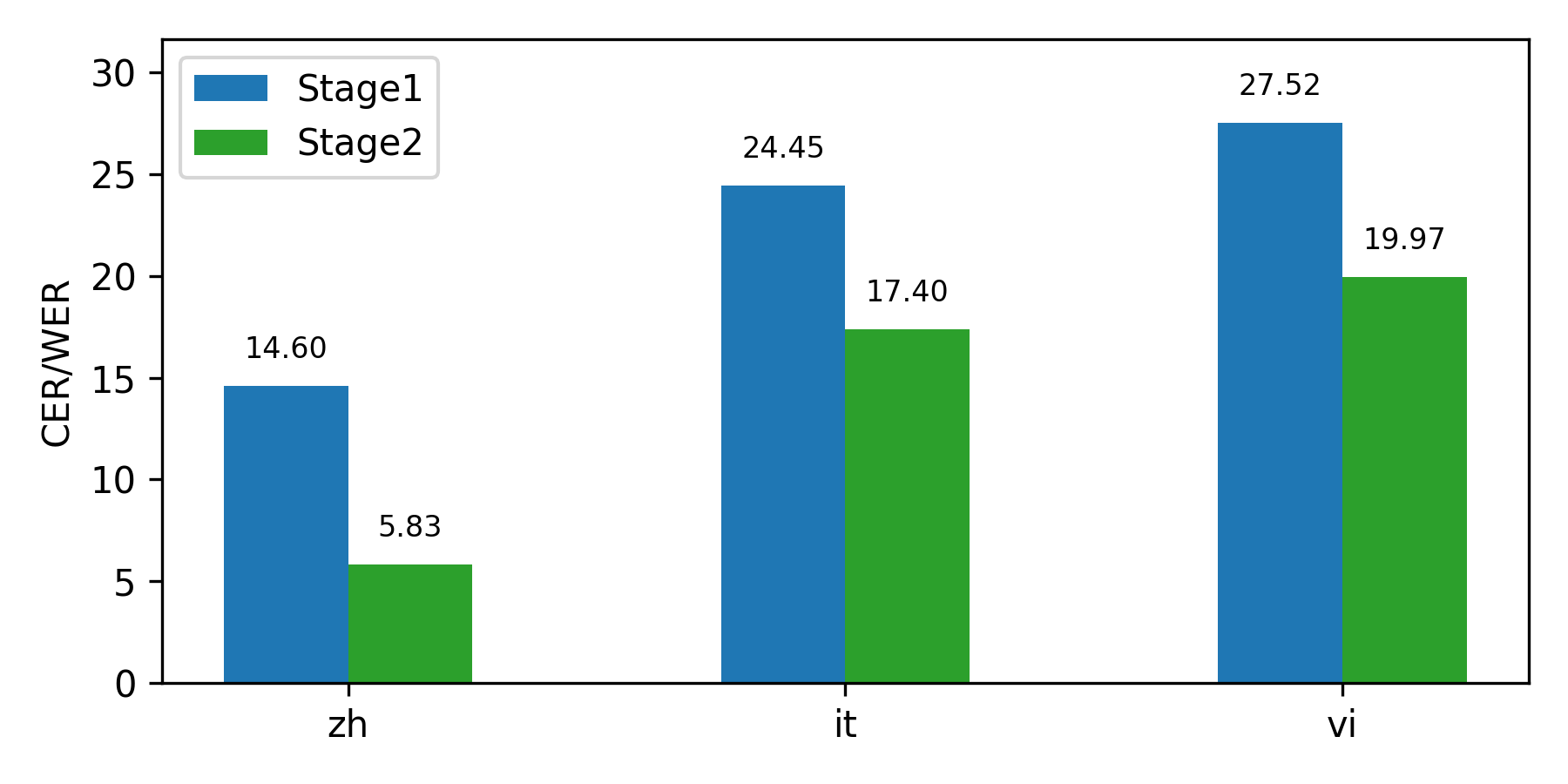}
    \caption{Performance of new language adaptation: CER for Chinese, WER for Vietnamese and Italian.}
    \label{fig:adaptation}
\end{figure}

We evaluate the adaptability of our approach to new languages by applying the model to speech data from previously unseen linguistic domains. In this experiment, we fine-tune the model in phase 2 using speech from three languages: Chinese (ZH), Italian (IT), and Vietnamese (VI). We use AISHELL-3 for Chinese \citep{aishell3}, the same dataset as \citep{vc-dat} for Vietnamese, and the Multilingual LibriSpeech (MLS) training subset for Italian \citep{mls}. We reserve a portion of each training set as a test set and pair 400 utterances for evaluation. To measure content intelligibility, we use language-specific ASR tools: FunASR \citep{funasr} with paraformer-zh \citep{paraformer} for Chinese, Chunkformer-large-vi \citep{chunkformer} for Vietnamese, and Whisper-large \citep{whisper} for Italian. Figure \ref{fig:adaptation} shows that Phase 2 fine-tuning boosts performance and enables language adaptation using only audio, without requiring labeled data. It also proves that our speaker-independent features still retain some accent or language information.

\subsection{Semantic Alignment of Synthetic Audio Pairs}
\label{sec:synthetic_align}

\begin{figure}[h]
  \centering
  \begin{subcaptionbox}{Cosine pairwise semantic similarity.\label{fig:proposed_alignment}}[0.85\linewidth]
    {\includegraphics[width=1\linewidth]{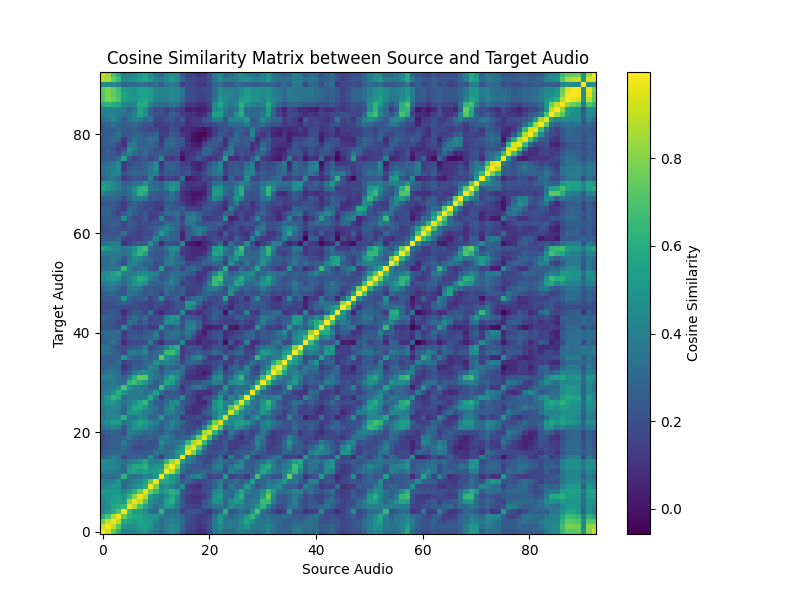}}
  \end{subcaptionbox}
  \hfill
  \begin{subcaptionbox}{Top-1 cosine similarity alignment path.\label{fig:top1_alignment}}[0.85\linewidth]
    {{\includegraphics[width=1\linewidth]{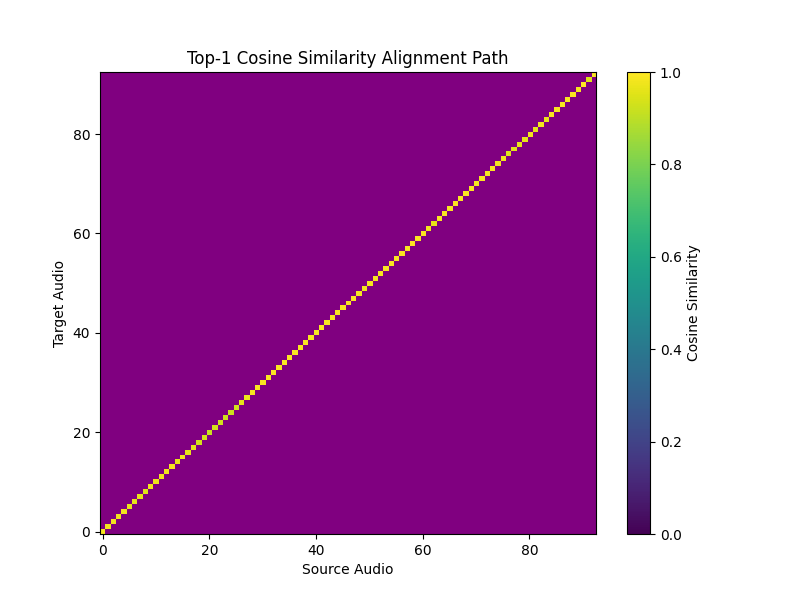}}}
  \end{subcaptionbox}
  \caption{Semantic alignment of source and target audio via synthetic data.}
  \label{fig:alignment}
\end{figure}
To evaluate the effectiveness of synthetic speech for input-output training, we examine the semantic alignment between the source and target audio generated. To assess semantic alignment quality, we extract semantic features with a pretrained HuBERT ASR model, as described in Section \ref{sec:evaluate-metrics}. We then compute the cosine similarity between all pairs of frames from the source and target audio, resulting in a pairwise similarity matrix. This matrix is visualized as a cosine similarity heatmap in Figure \ref{fig:proposed_alignment}.

The heatmap displays a clear diagonal of high similarity values, indicating strong frame-level alignment between the source and target audio. Furthermore, the top-1 cosine similarity alignment path, shown in Figure \ref{fig:top1_alignment}, lies precisely along the diagonal, confirming perfect alignment. These results demonstrate that the synthetic data input-output pairs are ideal training examples for voice conversion, enabling the model to learn effective one-to-one mapping.

\section{Conclusion}
We presented a robust voice conversion framework based on synthetic data and a two-phase training strategy. Our method enhances speaker similarity, speech quality, and content consistency, particularly in zero-shot scenarios with unseen target speakers. Experiments and ablation studies confirm the effectiveness of our approach and demonstrate its ability to prevent speaker information leakage from the source audio. Additionally, we showed that the model generalizes well to unseen languages without requiring labeled data, making it highly suitable for low-resource settings.

\section*{Limitations}
Although our model improves speaker similarity and content intelligibility, it still depends on access to a high-quality, well-labeled speech corpus to train the TTS system. Furthermore, the effectiveness of synthetic data generation and its influence on the performance of voice conversion across different TTS systems remain insufficiently explored. Therefore, in future work, we plan to investigate alternative TTS models to gain a deeper and more comprehensive understanding of their impact on overall system performance.

\section*{Ethics Statement}
Voice conversion technology raises ethical concerns because it can be misused to generate deepfake audio and illegally spoof someone’s identity without consent. Such applications risk enabling fraud, misinformation, and reputational harm, while also undermining public trust in authentic communication. Since voices are unique biometric identifiers, misuse poses serious threats to privacy and security. Ethical use requires informed consent, transparency, and robust safeguards, including reliable spoofing verification methods, to prevent malicious exploitation.


\bibliography{custom}

\appendix
\section{Appendix}

\label{sec:appendix}

\subsection{Preservation of Emotional Information}
\label{sec:emotion_keep}

\begin{figure}[h]
  \centering
  \begin{subcaptionbox}{Our Proposal\label{fig:proposed_emotion}}[1\linewidth]
    {\includegraphics[width=1\linewidth]{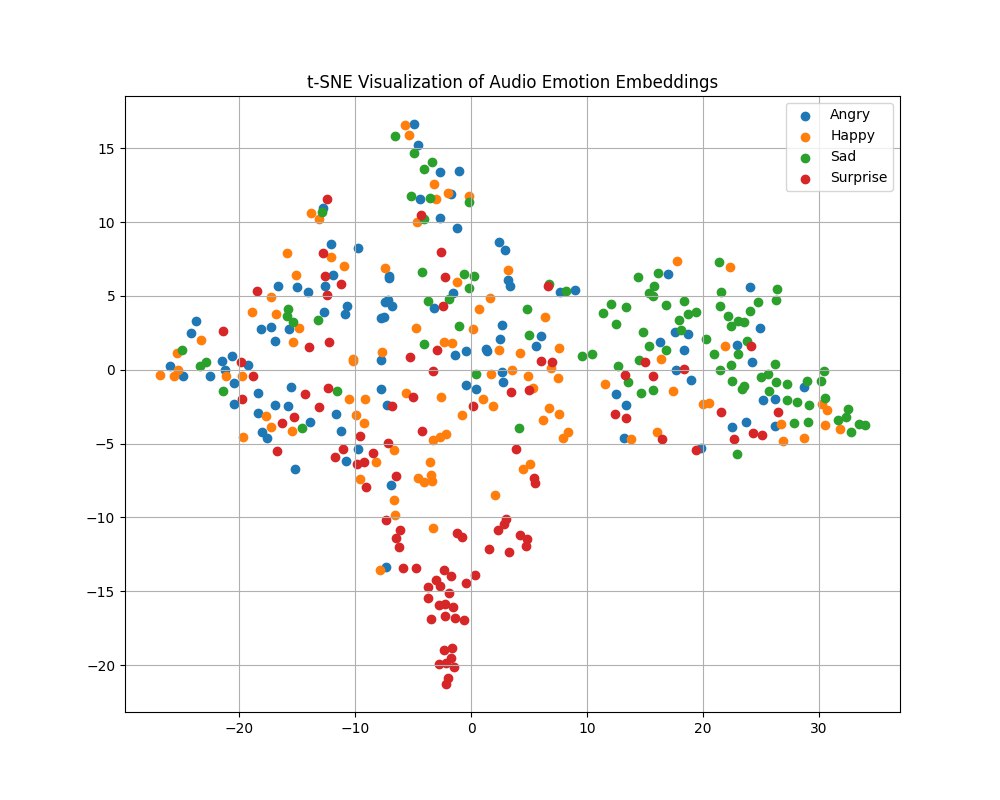}}
  \end{subcaptionbox}
  \hfill
  \begin{subcaptionbox}{FreeVC\label{fig:freevc_emotion}}[\linewidth]
    {{\includegraphics[width=1\linewidth]{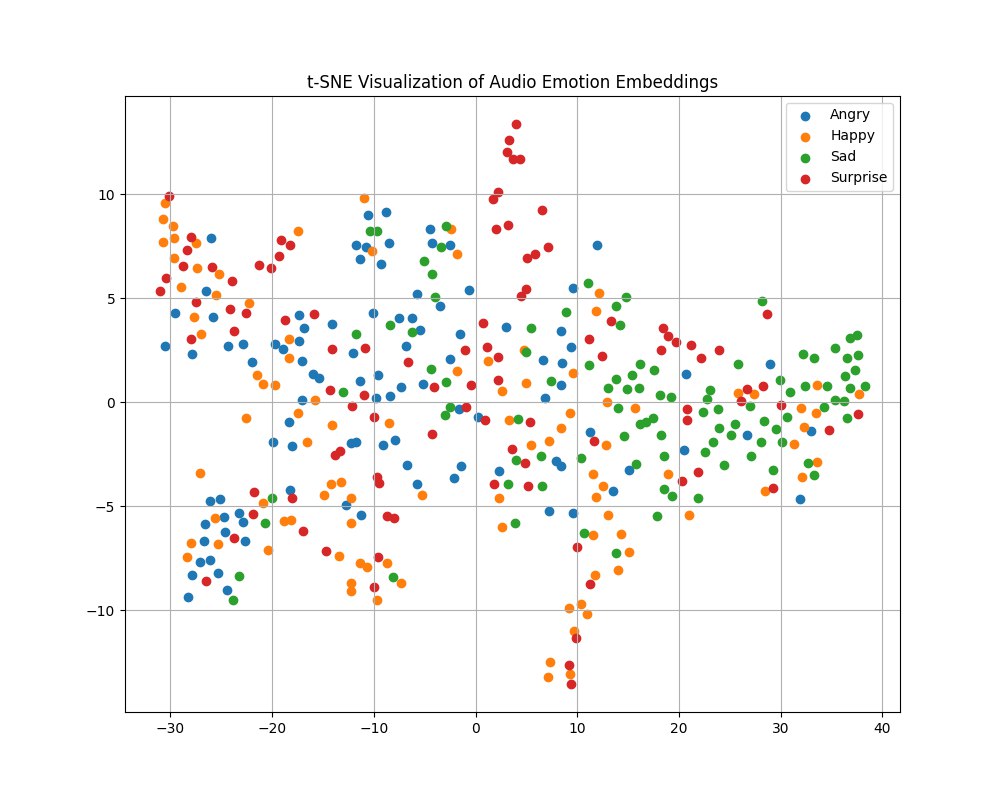}}}
  \end{subcaptionbox}   
  \caption{Emotion representation of converted audio using t-SNE.}
  \label{fig:emotion}
\end{figure}

To evaluate how well emotional information is preserved, we compare our proposed model with the baseline FreeVC, which shares the same backbone. For this experiment, we use the ESD dataset \cite{esd}, which contains emotional speech. We randomly sample 10 audio clips from each of the 10 speakers across 4 emotions, resulting in a total of 400 source audio samples. For the target speakers, we randomly select a speaker from the LibriSpeech test set. After performing voice conversion, we extract emotion embeddings from the converted audio using the emotion2vec\_plus\_large model \cite{emotion2vec} and visualize them using t-SNE. As shown in Figure \ref{fig:emotion}, our proposed model produces more distinct emotion clusters, such as sad and surprise, while FreeVC exhibits little to no clustering, indicating that our model better preserves emotional characteristics.

\begin{table}[t]
\centering
\begin{tabular}{|l|c|c|}
\hline
\textbf{Lang} & \textbf{Stage} & \textbf{CER/WER} \\ \hline
ZH & 1  & 16.59 \\ \hline
ZH & 2  & 7.76  \\ \hline
ZH & GT & 2.84  \\ \hline
IT & 1  & 28.93 \\ \hline
IT & 2  & 22.62 \\ \hline
IT & GT & 11.96 \\ \hline
VI & 1  & 11.10 \\ \hline
VI & 2  & 8.83  \\ \hline
VI & GT & 2.53  \\ \hline
\end{tabular}
\caption{CER/WER between converted audio and ground-truth transcriptions.}
\label{tab:native-like}
\end{table}

\subsection{Evaluation of Native-Like Qualities in Converted Speech}
\label{sub:native}

To assess how ``native-like'' the converted speech is, we report CER/WER between the converted audio and ground-truth transcriptions of the original voice. The datasets used for Chinese and Italian are the same as those described in Section~\ref{sec:adapt}. However, for Vietnamese, since the multi-speaker-vi dataset lacks text \cite{vc-dat}, we constructed a clean test set using utterances from the VIVOS dataset \cite{vivos} that do not overlap with the training data. Lower scores indicate higher intelligibility and a closer resemblance to native speech. Results are summarized in Table~\ref{tab:native-like}.

These results show clear gains in intelligibility after language adaptation, though still slightly below ground-truth levels. This indicates that the converted speech becomes substantially more native-like, while leaving room for further improvements to fully match natural speech.

\end{document}